# Design method of Temporary Horizontal Visibility Graph for information sources impact network

D.V. Lande and A.M. Soboliev
*Igor Sikorsky KPI, Kyiv, Ukraine*
*dwlande@gmail.com*

The absence of the efficient methods for the design of the information sources impact network does not allow defining the influence of the information sources on one another accurately and detecting the primary sources of the information spreading. The work represents a new approach - Temporary Horizontal Visibility Graph (THVG) method, which is based on the HVG algorithm modification. This method allows building the information sources impact networks and making the assumptions about the primary sources of information. Obtained method uses sources rating in data flow analysis. It is more effective than the usual Horizontal Visibility Graph design method in the following criteria: F-measure - by 7-9%, completeness - by 7-10% and accuracy - by 6-8% more effective.

Time series data analysis is essential to many fields of science and technology, in particular biology, physics, linguistics, economy, etc. It is worth mentioning that time series data analysis is used for forecasting, definition of the hidden periodicity and diagnostic tasks solving.

Consequently, complex time series studies are based on the usage of such numerical methods as static, fractal, correlation. For example, the calculation of Hurst exponent is determined by the indicators of the thematic data rate called fractal dimension. Namely thematic data flow studies prove the assumptions about the self-similarity and iteration of the information processes. Reprinting, citing, direct references cause similarity which shows up in proof statistical partitions and empirical regularities. Data bases self-similarity analysis can be considered as the approach to the range persistency definition - the definition of the possibility of the range to follow its previous trend.

Highly developed methods of the complex networks analysis, which are based on the "Visibility Graphs" [2], are used for the studies of the time series with nontrivial structure. Therewith Graphs are assigned to the time series according to the defined algorithm.

The design method of information sources interrelation network is used for the detection of the network of information sources and primary source time interrelations about the event of the defined theme.

The network designed in such a way reflects the sources connections, allows determining the most effective sources, making assumptions about the primary sources.

The research of the information source with the help of the InfoStream (the content monitoring system) has detected that the ranges of news amount pointings (when they are sequenced in time and correspond to the search theme) can be transformed into the Horizontal Visibility Graphs, in which not only the pointings correspond with the nodes, but the messages themselves. As a result, we design the Horizontal Visibility Graph from the obtained values of appearance time in the publication system (in horizontal direction) and the information sources rating (in vertical direction).

Therefor a range of publication nodes is marked on the horizontal centerline. Each of them corresponds with the nominal theme in order of appearance in the InfoStream system [3]. The values of the information sources rating which are corresponding with these publications are set on the vertical centerline. For designing the Horizontal Visibility Graph there must be the linkage between the nodes – the direct correlation present in cases when nodes are in the line of site. Namely, there is a linkage between the nodes if they can be connected with a horizontal line without piercing any other vertical line. This criterion is reflected on the pic. 1 and it can be depicted in the following way: two nodes (publications), e.g. $S^t$ and $S^{t+5}$, where $m = (t+5)$ are connected by the direct correlation.

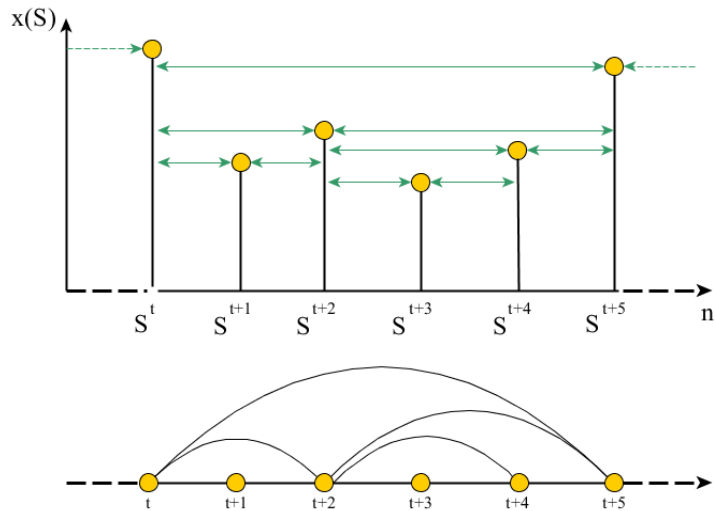

Pic. 1. The apparition between the nodes and linkage between the nodes.

On Pic. 1 the nodes with the messages $S^t$ та $S^{t+5}$ are neighboring for the message node $S^{t+2}$. Apart from it, $S^t$, which is the closest left-side informational message from $S^{t+2}$, and $S^{t+5}$ is the closest right-side message from $S^{t+2}$.

Thereupon the network received on the previous stage is compactified, and all the nodes from the specific information source, e.g. $S$, are combined in one node. All correlations of such nodes are combined as well. It is important to note that in such cases not more than 1 correlation is left between two nodes, multiple bonds are also excluded. In its turn, it means that the $S$ node degree during the calculations will not exceed the sum of powers $\sum_k A_k^n$.

As a result of passing the above mentioned stages we receive the new message network from the information sources – Horizontal Visibility Graph.

The method allows designing the message networks of the certain thematic field. Information sources, which have the high value act as descriptors (event triggers) and can be used for the popular events location and more specific data flow categorization from the information sources.

This method allowed using the messages published by the information sources and on the basis of these data creating the network of correlations between the information sources.

During the research process of the data flow from the InfoStream content monitoring, it was determined that when analyzing the incoming data, one can observe the information sources actively creating the one themed publications with small time period between publishing. Due to this fact, they get to content monitoring program, which has a negative impact on the influence determination and primary source, as the evaluation of the correlations between the information sources does not happen in full.

The Horizontal Visibility Graph method (HVG) was modified to calculate temporal correlations between the information sources and event primary sources in more detail. The new Temporary Horizontal Visibility Graph (THVG) method enables the construction of information source network interconnections, taking into account the possibility of a slight time order mismatch between the publications. A network is being created and it reflects the source links on a given topic, allows one to identify the most influential sources and make assumptions about the primary source.

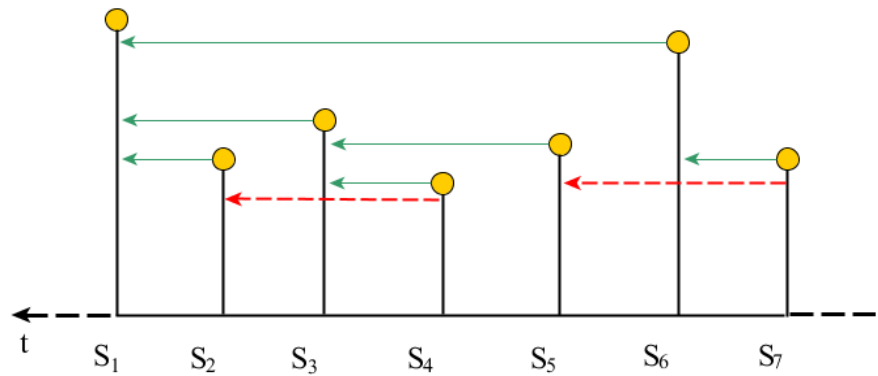

Pic. 2. Horizontal visibility of the messages from the information sources $S_i$ on the themed data flow using THVG method (where new links (which appeared due to method modification) are depicted by chain-dotted line).

The obtained Visibility Graph (THVG) design method from the time series $x(t_i)$ $(i=1,...,n)$ is that the horizontal centerline indicates the points corresponding to the time point $t_i$, from which segments of height equal to the value of the information source rating series at these points ($x(S)$) are built in the perpendicular direction. The nodes of Horizontal Visibility Graph are the outer vertices of the intervals. A connection between the vertices in THVG is considered to exist if the horizontal line drawn from one of the vertices intersects an interval of the other vertex. It should be noted that if a connection between the nodes at the first step happened, one must check if the previous node (with a value not more than $\tau$) is more significant than the current one. If so – an interval is drawn between these nodes.

Let us depict the THVG method in the form of mathematical formula (1):

$$a_{ij} = \begin{cases} 1, \text{ якщо } S_i > S_j \ \& \ j - i < \tau \\ 0, \text{ інше} \end{cases} \quad (1)$$

where $S_i$ – rating of the previous node as relating to the research node $S_j$;

$S_j$ – rating of the current research node;

$i$ – previous node as relating to the node $j$;

$j$ – current research node;

$\tau$ – a predetermined measure of distance from the current research node, an integer.

The results obtained using the formula (1) form the adjacency matrix (2):

$$A = \|a_{ij}\|_{i=\overline{1,N}, j=\overline{1,N}} \quad (2)$$

For visual representation let's show the connection between the graph nodes from Pic. 2 as an adjacency matrix.

$$A = \begin{pmatrix} 0 & 0 & 0 & 0 & 0 & 0 & 0 \\ 1 & 0 & 0 & 0 & 0 & 0 & 0 \\ 1 & 0 & 0 & 0 & 0 & 0 & 0 \\ 0 & 1 & 1 & 0 & 0 & 0 & 0 \\ 0 & 0 & 1 & 0 & 0 & 0 & 0 \\ 1 & 0 & 0 & 0 & 0 & 0 & 0 \\ 0 & 0 & 0 & 0 & 1 & 1 & 0 \end{pmatrix}, \quad (3)$$

The obtained adjacency matrix is necessary for the nodes influence determination and making the assumptions about the primary information source.

For graph density calculation we use the formula (4):

$$D = \frac{2v}{n(n-1)}, \quad (4)$$

where $n$ – the amount of nodes in the network;

$v$ – the amount of links between the nodes in the network.

Let's investigate the dependence of the network density obtained using the THVG method for different amount of nodes (Pic. 3)

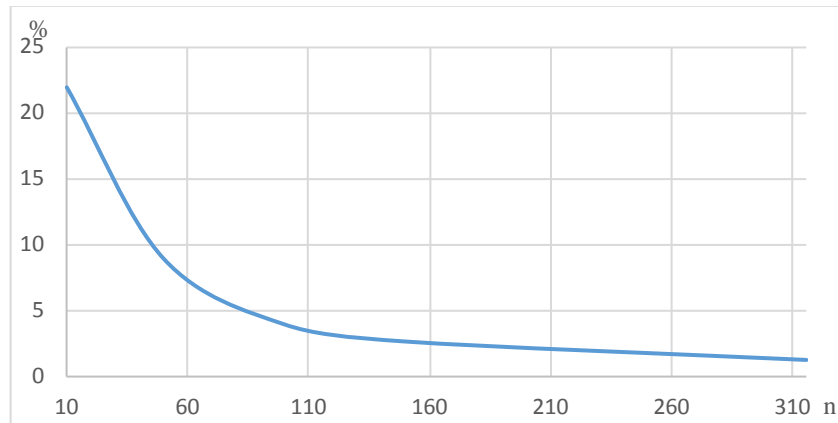

Pic. 3. The density graph of information sources influence network as relating to the amount of nodes *n*

As a result of the obtained network analysis using the THVG method the numerical regularity corresponding to formula (5) with an accuracy of 0.947 was found.

$$D == -6{,}283\ln(x) + 34{,}924, \qquad (5)$$

This implies that as the number of nodes increases, the density of the obtained network gradually decreases. This allows the analysis of an incompletely connected network which does not significantly increase the complexity of the study of connections in networks with a great amount of nodes.

The mentioned approach is based on the assumption that online media sources are pre-ranked by the volumes of publications and rates of publishing, based on the continuous observation. On the basis of a series of publications from the data flow, each of which corresponds with the information source where the information was published and the time of its publication on the network, the horizontal visibility graph is built. The "gaze direction" is set on the future. In this way the source of information is linked to another, more rated one, if one exists and published the information earlier. The network thus designed reflects the connections of the sources on the given subject, allows to determine the most influential sources among them, to make assumptions about the primary information source about an event.

To determine the performance of the THVG visibility graph method compared to the conventional HVG visibility graph method, in the process of identifying temporal relationships between information sources an analysis of the incoming information flow from the InfoStream content monitoring system was performed resulting in:
- The modified method analyzed the input data and based on the obtained analysis identified 22% more connections between information sources;
- The diameter of the resulting network using the modified method is 11 (if comparing with the conventional method it is only 9);
- The density of the resulting network increased from 0.005 to 0.007;
- The average degree of network using the modified method increased from 1.492 to 2.209.

An assessment has been carried out on the results of determining the influence of the information sources nodes from the data flow of online media sources content monitoring using the modified method of constructing the visibility graph. Accuracy, completeness and overall F-measure scores were evaluated. The results are shown below.

| Метод | Повнота (Recall) | Точність (precision) | F–міра |
|---|---|---|---|
| THVG | 96,8% | 71,4% | 82,2% |
| HVG | 87,3% | 63,5% | 73,5% |

The assessment was conducted on the basis of information sources from the news content monitoring service InfoStream with publications on the topic "cybersecurity" from information sources accumulated over 3 calendar months with a total activity of more than 5000 active sources. To accomplish this task the expert method required a much larger amount of time. In doing so, the accomplishment becomes possible either for a very large staff or by using the proposed modified method of constructing an HVG graph. Thus it becomes obvious that the proposed method is much faster than the expert determination of the nodes impact and the assumption of the primary source. The accuracy and completeness are higher than the available methods of automated research of such data.

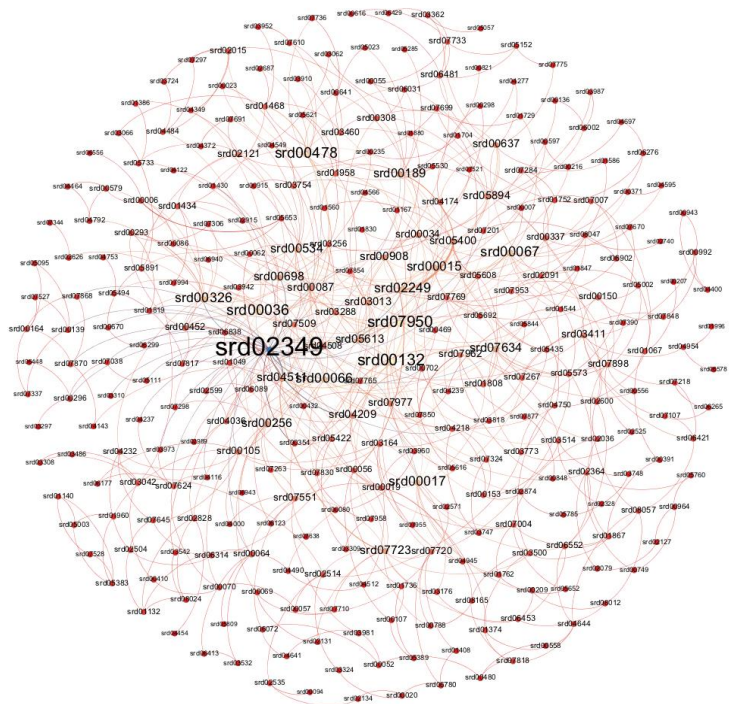

Pic. 4 The obtained links between information sources using the modified cybersecurity design method of HVG visibility graph

**Conclusion**

The paper proposes a new design method of information sources mutual impact network Temporary Horizontal Visibility Graph (THVG) which is a modified method of horizontal visibility graph (HVG). The method allows to determine the influential information sources taking into account the time dependence of information influences, their relevance and "aging" of information. The information sources network designed in such a way can dynamically change. Thus, new nodes – information sources – can appear. This approach allows the use of source ratings for the data flow analysis and is more effective than the conventional method of constructing the Horizontal Visibility Graph (HVG) in terms of: F-measure - by 7-9%, completeness – by 7-10%, accuracy – by 6-8%.